\pdfoutput=1
\documentclass[pra,twocolumn,superscriptaddress,nofootinbib]{revtex4}

\usepackage{amsmath,amsfonts,amssymb,amstext}
\usepackage[dvipdfm]{graphicx}
\usepackage[colorlinks,linkcolor=blue]{hyperref}
\usepackage{multirow}

\newcommand{\bra}[1]{\ensuremath{\langle #1 \vert}}
\newcommand{\ket}[1]{\ensuremath{\vert #1 \rangle}}
\newcommand{\braket}[2]{\ensuremath{\langle #1 \vert #2 \rangle}}
\newcommand{\be}{\begin{equation}}
\newcommand{\ee}{\end{equation}}
\newcommand{\bea}{\begin{eqnarray}}
\newcommand{\eea}{\end{eqnarray}}
\newcommand{\bF}{\begin{figure}}
\newcommand{\eF}{\end{figure}}

\newcommand{\bi}{\begin{itemize}}
\newcommand{\ei}{\end{itemize}}

\newcommand{\mbf}[1]{\mathbf{#1}}

\begin{document}
\title{Constructing General Unitary Maps from State Preparations}
\date{\today}

\author{Seth T. Merkel}
\affiliation{Department of Physics and Astronomy, University of New Mexico, Albuquerque, NM, 87131, USA}

\author{Gavin Brennen}
\affiliation{Physics Department, Macquarie University, NSW, 2109, Australia}

\author{Poul S. Jessen}
\affiliation{College of Optical Sciences, University of Arizona, Tucson, Arizona 85721, USA}

\author{Ivan H. Deutsch}
\affiliation{Department of Physics and Astronomy, University of New Mexico, Albuquerque, NM, 87131, USA}

\begin{abstract}
We present an efficient algorithm for generating unitary maps on a $d$-dimensional Hilbert space from a time-dependent Hamiltonian through a combination of stochastic searches and geometric construction.  The protocol is based on the eigen-decomposition of the map.  A unitary matrix can be implemented by sequentially mapping each eigenvector to a fiducial state, imprinting the eigenphase on that state, and mapping it back to the eigenvector.  This requires the design of only $d$ state-to-state maps generated by control waveforms that are efficiently found by a gradient search with computational resources that scale polynomially in $d$.  In contrast, the complexity of a stochastic search for a single waveform that simultaneously acts as desired on all eigenvectors scales exponentially in $d$.   We extend this construction to design maps on an $n$-dimensional subspace of the Hilbert space using only $n$ stochastic searches.  Additionally, we show how these techniques can be used to control atomic spins in the ground electronic hyperfine manifold of alkali atoms in order to implement general qudit logic gates as well to perform a simple form of error correction on an embedded qubit.   
\end{abstract}

\maketitle

\section{Introduction}
The goal of quantum control is to implement a nontrivial dynamical map on a quantum system as a means to achieve a desired task.  Historically, the major developments in quantum control protocols have been motivated by applications in physical chemistry whereby shaped laser pulses excite molecular vibrations and rotations \cite{shapiro86,judson92}, and in nuclear magnetic resonance whereby shaped rf pulses cause desired spin rotations in magnetic-resonance-imaging \cite{khaneja01,cory05,vandersypen04}.  More recently, quantum control theory has been considered in the development of quantum information processors in order to tackle the challenges of extreme precision and robustness to noise and environmental perturbations \cite{viola99, vandersypen04, khaneja05a, grace07, chiara08}. Such quantum processors are being explored on a wide variety of platforms ranging from optics and atomic systems, to semiconductors, and superconductors.  The design of new protocols for quantum control can thus impact a wide spectrum applications.

The simplest approach to quantum control is open-loop unitary evolution.  In this protocol, the system of interest is governed by a Hamiltonian that is a functional of a set of time-dependent classical ``control waveforms", $H\left[\mbf{B}(t) \right]$.  Through an appropriate choice of  $\mbf{B}(t)$, the goal to reach a desired solution to the time-dependent Schr\"{o}dinger equation at time $T$, formally expressed as a time-ordered exponential, $U(T)=\mathcal{T}\left(\exp \left\{-i \int_0^T H\left[ \mbf{B}(t')  \right] dt' \right\} \right)  $.  The system is said to be ``controllable" if for any $W$ in the space of unitary maps on the Hilbert space of interest, there exists a set of control waveforms that such that $U(T)=W$ for some time $T$.  Control theorists have long known the conditions on  $H\left[ \mbf{B} (t) \right]$ such that the system is controllable in principle, but a construction for specifying the desired waveforms is generally unknown.  The goal of this paper to to provide such a construction for a wide class of quantum systems.  We restrict our attention to Hilbert spaces of finite dimension $d$. 

Two classes of quantum control problems have been primarily considered: state-preparation and full unitary maps.  In state-preparation, the goal is to map a known fiducial initial quantum state $\ket{\psi_i}$ to an arbitrary final state $\ket{\psi_f}$.  This requires specification of only one column of the unitary matrix, i.e. the vector $U(T) \ket{\psi_i}$, as compared with the full unitary map, which requires specification of all $d$ orthonormal column vectors.  The contrast between these tasks is reflected in the complexity of numerical searches for the desired waveforms.  Optimal control theory provides a framework for carrying out such searches \cite{werschnik07}.  An objective function $J$ is defined for the task at hand, e.g., $J[\mbf{B}(t)] = \bra{\psi_f} U(T) \ket{\psi_i}$ for state preparation or $J[\mbf{B}(t)] = \text{Tr}\left( W^{\dagger} U(T) \right)$ for full unitary mapping.  The optimal controls are the maxima of these objective functions.   

In series of papers, Rabitz and coworkers introduced the concept of the ``control landscape" \cite{rabitz04,shen06,hsieh08, moore08}.  By discretizing the control functions (e.g., by sampling at discrete times), one can treat the objective as a smooth function whose domain is a finite set of control variables.  The topology of this resulting hypersurface governs the complexity with which numerical search algorithms can find optimal solutions.  In the case of state preparation, Rabitz {\em et al.} showed that for open-loop unitary control, the control landscape has an extremely favorable topology \cite{rabitz04}.  Given a closed-system open-loop Hamiltonian evolution for sufficient time $T$, all critical points (i.e. those values of the control parameters where $\delta J =0$) are either unit fidelity or zero fidelity; there are neither local optima nor saddle points.  Furthermore,  the surface has a gradual slope as one moves towards the optimal points, and there are an infinite number of optimal solutions connected on a submanifold with large dimension, $N_c-2d+2$, where $N_c \ge d^2-1$ is the number of control variables defining the dimension of the overall objective-function hypersurface \cite{shen06}.  The lack of false suboptimal critical points, the gentle slope, and the flat region near a maximum, all enable efficient search algorithms that yield fairly robust optimal control waveforms based on a simple gradient ascent algorithm from a random seed.  A collection of random seeds yield a collection of possible solutions that can then be further tested for robustness to decoherence and noise.  

In contrast, the control landscape for full unitary control is less favorable \cite{hsieh08}.  For $SU(d)$ matrices, there are $d$ critical values of the objective function.  Of these, there is one optimal solution with unit fidelity, an isolated point in the control landscape.  The remaining $d-1$ suboptimal points are saddles.  While the lack of local maxima may suggest numerical optimization might still be an efficient search strategy, empirical studies show otherwise \cite{moore08}. Whereas state preparation search routines converge in a number of iterations that is essentially independent of $d$, the resources necessary for optimization algorithms to converge on the full unitary control landscape grow {\em exponentially} with $d$.  Brute force search is thus a very poor strategy for full unitary control on all but the smallest dimensional Hilbert spaces.

Explicit constructions for full unitary control have been established in special cases where the form of the Hamiltonian allows.  For example, Khaneja {\em et al.} showed that the problem of generating unitary matrices on a system of weakly coupled qubits can be reduced to the solution of a geodesic equation \cite{khaneja01}.  Brennen {\em et al.} showed that by considering controls on overlapping 2-$d$ subspaces it is possible to create arbitrary controls through Givens rotations \cite{brennen05}.  Such constructive procedures are less computationally intensive than their random search counterparts, and moreover, yield control fields that are more physically intuitive.  They are, however, restricted to control systems with particular structures and are not applicable in more generic cases.

In this paper we develop a hybrid protocol for full unitary control that combines efficient numerical search procedures with constructive algorithms, applicable for any finite dimensional Hilbert space with minor restrictions, thereby extending the work of  Luy {\em et al.} \cite{luy05}. We leverage off of the efficiency of numerical searches for waveforms that generate a desired state mapping.  Our procedure requires only $d$ stochastic searches and the length of the resulting control sequence is approximately $2d$ times the time of a single state preparation.  Our work is analogous to that of \cite{moore08}, where state-preparation provides a good starting point for iterative searches.  Our procedure, however, is fully constructive and deterministic once appropriate state mappings are found.
 
The remainder of this paper is structured as follows.  In Sec.~\ref{sec:construct} we present our hybrid protocol for constructing general unitary maps by combining efficient numerical searches with a deterministic algorithm.   In addition to unitary maps on the full Hilbert space, this scheme allows us to construct maps on a subspace with a complexity that scales as the dimension of that space.  Finally, in Sec.~\ref{sec:examples}, we apply our unitary matrix construction to control the large manifold of magnetic sublevels in the ground electric states of an alkali atom (e.g. $^{133}$Cs) \cite{merkel08}.   We show how to construct a set of unitary matrices on $SU(d)$ that are often considered as qudit logic gates in a fault-tolerant protocol.  In addition, we apply our construction for subspace mapping to encode logical qubits in our qudit, and simulate an error correcting code that protects against magnetic field fluctuations.

\section{Unitary Construction}\label{sec:construct}

In this section we define an efficient protocol for constructing arbitrary unitary maps based on state preparation.  Any unitary matrix has an eigen-decomposition, 
\be
U = \sum_j e^{-i \lambda_j} \ket{\phi_j}\bra{\phi_j}= \prod_j e^{-i \lambda_j  \ket{\phi_j}\bra{\phi_j}},  
\ee 
where in the second form we expressed $U$ as a product of commuting unitary evolutions by moving the projectors into the exponential.  A general unitary map can be thus be constructed from $d$ propagators of the form $\exp\{-i \lambda_j  \ket{\phi_j}\bra{\phi_j}\}$, one for each eigenvalue/eigenvector pair.   These unitary propagators can now be constructed using state mappings.  We begin by noting that there exists some $V_j \in SU(d)$ that satisfies 
\be
e^{-i \lambda_j  \ket{\phi_j}\bra{\phi_j}} = e^{-i \lambda_j  V_j^{\dagger} \ket{0}\bra{0} V_j} = V_j^{\dagger} e^{-i \lambda_j  \ket{0}\bra{0}} V_j,
\ee 
where $\ket{0}$ is a fixed ``fiducial state".  The sole requirement on $V_j$ is that $|\bra{0} V_j \ket{\phi_j}|^2 = 1$, i.e., it must be a mapping from $\ket{\phi_j}$ to $\ket{0}$.  Therefore, we can create the unitary propagator $\exp\{-i \lambda_j  \ket{\phi_j}\bra{\phi_j}\}$ by using  a state preparation to map the eigenvector of $U$, $\ket{\phi_j}$, onto the fiducial state $\ket{0}$, applying the correct phase shift, and finally mapping the fiducial state back to the eigenvector with the time-reversed state preparation.  A general unitary map is thus constructed via the sequence,
\be
U=V_d^{\dagger} e^{-i \lambda_d  \ket{0}\bra{0}} V_d \ldots V_{2}^{\dagger} e^{-i \lambda_{2}  \ket{0}\bra{0}} V_{2}V_1^{\dagger} e^{-i \lambda_1  \ket{0}\bra{0}} V_1  .
\ee  
Each of the propagators $V_j$ is specified by a control waveform that generates a desired state mapping.  One can efficiently find such control fields based on a numerical search that employs a simple gradient search algorithm, as described above.  To generate an arbitrary element of $SU(d)$, we require at most $d$ such searches. Moreover, the full construction consists of $2d$ state preparations interleaved with $d$ applications of the phase Hamiltonian, leading to an evolution that is only of order $d$ times longer than a state mapping evolution.              

This construction places only two requirements on the Hamiltonian in addition to controllability.  Firstly, the dynamics must be reversible such that if we can generate the unitary evolution $V_j$, we can trivially generate the unitary $V_j^{\dagger}$ by time-reversing the control fields.  Note that this is not the same as finding a state preparation that goes in the opposite direction, $\ket{0} \rightarrow \ket{\phi_j}$; there are many unitary propagators that map $\ket{0}\rightarrow \ket{\phi_j}$, so it is unlikely to find the unique operator $V_j^{\dagger}$ from a stochastic search.  Secondly, we require access to a control Hamiltonian that applies an arbitrary phase to one particular fiducial state $\ket{0}$ relative to all of the remaining states in the Hilbert space, $\exp\{-i \lambda_j  \ket{0}\bra{0}\}$.  This latter requirement is the most restrictive, but can be implemented in a wide variety of systems.  An example is discussed in Sec.~\ref{sec:examples}.

\subsection{Subspace Maps}\label{sec:subspace}
We have so far considered two kinds of maps on our $d$-dimensional Hilbert space $\mathcal{H}$: $d \times d$ unitary matrices and state-to-state maps.  The former corresponds to a map $U : \mathcal{H} \rightarrow \mathcal{H}$, while the latter specifies a map on a one-dimensional space.  Intermediate cases are also important.  In particular, we are often interested in unitary maps that take subspace $\mathcal{A}$ of arbitrary dimension $n$ to subspace $\mathcal{B}$,  according to $T: \mathcal{A} \rightarrow \mathcal{B}$.  Examples include the encoding of a logical qubit into a large dimensional Hilbert space $( \mathcal{A} \neq \mathcal{B})$ and a logical gate on encoded quantum information $( \mathcal{A} =\mathcal{B})$. Above we showed that the design of a fully specified unitary matrix required search for $d$ waveforms that define $d$ state preparations (trivially a state mapping requires one such search).  We show here how unitary maps on subspaces of dimension $n$ can be constructed from exactly $n$ such numerical solutions.

Formally, a unitary map between two subspaces  $ \mathcal{A}$ and $\mathcal{B}$  of dimension $n$ is defined as a map between between their orthonormal bases $\{\ket{a_i}\}$ and $\{\ket{b_i}\}$,
\be
T_n\left(\mathcal{A} \rightarrow \mathcal{B}\right) =  \sum_{i=1}^n \ket{b_i}\bra{a_i} \oplus U_{\perp},
\label{eq:desiredmap}
\ee
where $U_{\perp}$ is an arbitrary unitary map on the orthogonal complement $ \mathcal{A}_{\perp}$ whose dimension is $d-n$.  State preparation is the case $n=1$; a full unitary matrix is specified when $n=d$.  Clearly for $n \neq d$ the map is not unique, with implications for the control landscape and the simplicity of numerical searches described above.  As a first na\"{i}ve construction of $T(\mathcal{A} \rightarrow \mathcal{B})$, one might consider a sequence of one-dimensional state mappings, 
\be
T_n\left(\mathcal{A} \rightarrow \mathcal{B}\right) \stackrel{?}{=} \prod_{i=1}^n T_1\left(\ket{a_i} \rightarrow \ket{b_i}\right). 
\label{eq:bad_map}
\ee
This does not, however, yield the desired subspace map because each state mapping acts also on the orthogonal complement, so, e.g. $\ket{b_1}$ is affected by $T_1\left(\ket{a_2} \rightarrow \ket{b_2}\right)$ and subsequent maps will move formerly correct basis vectors to arbitrary vectors in the orthogonal component.  We can resolve this problem by instead constructing subspace maps as a series all well-chosen rotations that maintain proper orthogonality conditions.

To construct the necessary unitary operators, we make use of the tools described above: arbitrary state mapping based on an efficient waveform optimization and phase imprinting on a fiducial state.  With these, we define the unitary map between unit vectors $\ket{a}$ and $\ket{b}$,
\be
S\left( \ket{a} ,\ket{b} \right) \equiv e^{-i \pi \ket{\phi}\bra{\phi}} = \hat{I} - 2 \ket{\phi}\bra{\phi}.
\ee
Here $\ket{\phi}=N(\ket{a} - \ket{b})$, where we have chosen the phases such that $\braket{b}{a}$ is real and positive, and $1/N^{2} \equiv 2\left(1-\braket{b}{a}\right)$ is the normalization.  This unitary operator has the following interpretation.  In the two-dimensional subspace spanned by $\ket{a}$ and $\ket{b}$, $S$ is a $\pi$-rotation that maps $S\ket{a} = \ket{b}$.  In contrast to the state preparation map, Eq.~(\ref{eq:desiredmap}) with $n=1$, this map acts as the {\em identity} on the orthogonal complement to the space.  This property is critical for the desired application.

With these 2D primitives in hand, we can construct the subspace map according to the prescription,
\be
T_n(\mathcal{A} \rightarrow \mathcal{B}) =s_n\ldots s_2 s_1,
\ee
where $s_k \equiv S\left( \ket{\tilde{a}_k} ,\ket{b_k} \right)$ and
\be
\ket{\tilde{a}_j}  \equiv s_{j-1}\ldots s_2 s_1 \ket{a_j}. 
\ee
This sequence does the job because each successive rotation leaves previously mapped basis vectors unchanged.  To see this, we must show that at step $j$, the basis vectors $\{ \ket{b_1},\ket{b_2}, \ldots, \ket{b_{j-1}} \}$ are unchanged by $s_j$.  This will be true when this set is orthogonal to the vectors $\ket{\tilde{a}_j}$ and $\ket{b_j}$.  Orthogonality to $\ket{b_j}$ is trivial since the basis vectors of $\mathcal{B}$ are orthonormal.  We must thus prove, $\braket{\tilde{a}_j}{b_k}=0$,  $\forall j>k$.  We can do this by induction.  For an arbitrary $k$, assume the conjecture is true for all $j$ such that $j_0 \geq j>k$, and thus $s_j \ket{b_k} = \ket{b_k}$ up to $j=j_0$.  This implies that $\braket{\tilde{a}_{j_0+1}}{b_k} = 0$ since,
\bea
\braket{\tilde{a}_{j_0+1}}{b_k} &=& \bra{a_{j_0+1}} s^{\dagger}_1 \ldots s^{\dagger}_k s^{\dagger}_{k+1} \ldots s^{\dagger}_{j_0} \ket{b_k} \nonumber\\ 
&=& \bra{a_{j_0+1}} s^{\dagger}_1 \ldots s^{\dagger}_k  \ket{b_k} \nonumber\\ 
&=& \braket{a_{j_0+1}}{a_k}=0.
\eea
To complete our proof by induction, we must show that for any $k$, the conjecture is true for $j=k+1$.  This follows since, 
\bea
\braket{\tilde{a}_{k+1}}{b_k} &=& \bra{a_k+1}s^{\dagger}_1 s^{\dagger}_2 \ldots s^{\dagger}_k \ket{b_k} \nonumber\\
&=& \braket{a_{k+1}}{a_k} =0.
\eea
With this protocol we can construct unitary maps on a subspace of dimension $n$ with optimized waveforms that corresponded to exactly $n$  prescribed state preparations.  In the following section we apply these tools to qudit manipulations in atomic systems.

\section{Applications to atomic spin control}\label{sec:examples}

\begin{figure}[t]
\begin{center}
\includegraphics[width=8cm,clip]{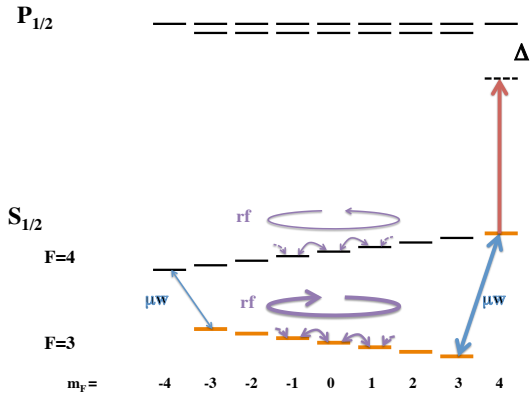}
\caption{The hyperfine structure of $^{133}$Cs in  the 6S$_{1/2}$ ground state.  Microwaves (blue) and rf magnetic fields (purple) provide controllable dynamics on the 16-dimensional Hilbert space.  A detuned laser light shift (red) can be used to create a relative phase between the $F=4$ and $F=3$ manifolds. By considering controls on the subspace of the orange states we recover a system that satisfies the criteria proposed in Sec.~\ref{sec:construct}. }
\label{F:levels}
\end{center}
\end{figure}

In this section, we apply our results to the control of the ground-electronic manifold of magnetic sublevels in alkali atoms.  Atomic spins are natural carriers of quantum coherence for use in various quantum information processing applications \cite{molmer08,polzik04,kuzmich05,kimble08}.  In previous work we showed that the full ground-electronic subspace of coupled electron and nuclear spin can be rapidly controlled through combinations of static, radio, and microwave-frequency ac-magentic fields, with negligible decoherence \cite{merkel08}.  A schematic for the specific case of $^{133}$Cs, with nuclear spin $I=7/2$ and two ground-electronic hyperfine manifolds with total angular momentum $F=3$ and $F=4$, is shown in Fig.~\ref{F:levels}.  A static bias magnetic field breaks the degeneracy and specifies an rf-resonance frequency by the Zeeman splitting in a given manifold.  Control of the amplitude and phase of the rf-magnetic fields oscillating in two spatial directions allows one to independently rotate these two manifolds.  Resonant microwaves can be used to excite transitions between $F=3$ and $F=4$, driving coherent $SU(2)$ rotations between two magnetic sublevels, as specified by a given (nondegenerate) transition frequency.  Such controls together can be used to generate an arbitrary unitary transformation on the $d=2(2I+1)=16$ dimensional Hilbert space.  In our previous work we showed how we could design state-preparation mappings through simple gradient searches \cite{merkel08}.  In the present work we show how we can employ this tool to design more general unitary maps based on the protocol of Sec.~\ref{sec:construct}.

In addition to an efficient method for designing and implementing state-to-state mappings, our protocol places certain requirements on the available control tools. Firstly, the system dynamics must be reversible so that we can trivially invert a state mapping.  This is easily achieved through phase control.  Secondly, we require phase imprinting on a single fiducial state.  While this cannot be accomplished using solely microwave and rf-control, by introducing an excited electronic manifold, an off-resonant laser-induced light-shift can achieve this goal.  We restrict our system to one spin manifold (here the $F=3$, but in principle either will do) and a single state from $F=4$ manifold, e.g. $\ket{F=4,m=4}$, which acts as the fiducial state.  By detuning far compared to the excited state line width of  ~5 MHz, but close compared to the ground-state hyperfine splitting of ~10 GHz, we imprint a light shift solely on the $\ket{F=4,m=4}$ state with negligible decoherence. Using rf-magnetic fields to perform rotations in the $F=3$ manifold, and microwaves to couple to the fiducial state, we obtain  controllable and reversible dynamics.  Note that we may include the fiducial state in our Hilbert space, for a total of 8 sublevels, or treat it solely as an auxiliary state and restrict the Hilbert space to the 7-dimensional $F=3$ manifold.

\subsection{Constructing qudit unitary gates}
The standard paradigm for quantum information employs two-level systems -- qubits -- in order to implement binary quantum-logic based on $SU(2)$ transformations.  Extensions beyond binary encodings in $d>2$ system -- qudits -- based of $SU(d)$ transformations have also been studied and may yield advantages in some circumstances \cite{gottesman99,brennen05, grace06}.  Of particular importance for fault-tolerant operation is implementation of these transformations through a finite set of ``universal gates".  Our goal here is to show how important members of the universal gate set can be implemented using our protocol.

In choosing a universal gate set appropriate for error correction, it is natural to consider generalizations of the Pauli matrices $X$ and $Z$ which generate $SU(2)$.  The generalized discrete Pauli operators for $SU(d)$ are defined
\bea
X \ket{j} &=& \ket{j \oplus 1}\nonumber\\
Z \ket{j} &=& \omega^j \ket{j}.
\eea
Here $\oplus$ refers to addition modulo $d$ and $\omega$ is the primitive $d$th root of unity, $\omega =\exp \{i 2 \pi /d\}$.  By considering the commutation relation of $X$ and $Z$, the remaining generalized Pauli operators have the form $\omega^l X^j Z^k$, defining the elements of Pauli group for one qudit (up to a phase).  This group is a discrete (finite dimensional) generalization of the Weyl-Heisenberg group of displacements on phase space.  

Another important group of unitary matrices in the theory of quantum error correction is the Clifford group, given its relationship to stabilizer codes \cite{gottesman99}.  These group elements map the Pauli group back to itself under conjugation.  Expressed in terms of their conjugacy action on $X$ and $Z$, the generators of the Clifford group for single qudits are
\bea
HXH^{\dagger} = Z, &\quad &HZH^{\dagger} = X^{-1} \\ 
SXS^{\dagger} = XZ, &\quad &SZS^{\dagger} = Z \\ 
G_a X G_a^{\dagger} = X^a, &\quad &G_a ZG_a^{\dagger} = Z^{a^{-1}}\nonumber\\ &&\textrm{when gcd}(a, d)=1
\eea
$H$ and $S$ are direct generalization of the Haddamard and phase-gates familiar for qubits \cite{nielsen2000}.  The $d$-dimensional $H$  is the discrete Fourier transform
\be
H\ket{j} = \frac{1}{\sqrt{d}}\sum_k \omega^{jk}\ket{k}
\ee           
and $S$ is a nonlinear phase gate
\bea
S\ket{j} = \omega^{j(j-1)/2}\ket{j} \quad j~\textrm{odd},\\
S\ket{j} = \omega^{j^2/2}\ket{j} \quad j~\textrm{even}.
\eea
The operator $G_a$ is a scalar multiplication operator with no analog in the standard Clifford group on qubits, defined by
\be
G_a \ket{j} = \ket{a j},
\ee
where the multiplication is modulo $d$.  The only such multiplication operator for 2-level systems is the identity operator.

While both the generalized Pauli and Clifford groups have utility in quantum computing, it is clear from their descriptions that unlike their qubit $SU(2)$ counterparts, these unitary matrices do not arise naturally as the time evolution operators governed by typical Hamiltonians.  This fact is not relevant to our unitary construction, which requires only knowledge of the operators' eigenvectors and eigenvalues.  Using the time-dependent Hamiltonian dynamics with couplings illustrated in Fig.\ref{F:levels} we have engineered control fields to create the generators of both the Pauli and Clifford groups acting on the 7-dimension $F=3$ hyperfine manifold.  The duration of  waveforms is approximately 1.5 ms, which is significantly shorter than the decoherence time of the system.  In principle, the durations of these waveforms could be decreased by an order of magnitude or more by using more powerful control fields.  Our objective function for creating a desired unitary $W$ is the trace distance $J[W] = Tr\left(W^{\dagger} U \right)$, where $U$ is the unitary matrix generated by our control waveforms.  Based on our protocol, employing state mappings that have fidelities of ~0.99, our construction yields unitary maps that reach their targets with fidelities of $J[Z] = 0.9866$, $J[X] = 0.9872$, $J[H] = 0.9854$, $J[S] = 0.9892$ and $J[G_3] = 0.9801$.

\begin{figure}[t]
\begin{center}
\includegraphics[width=8cm,clip]{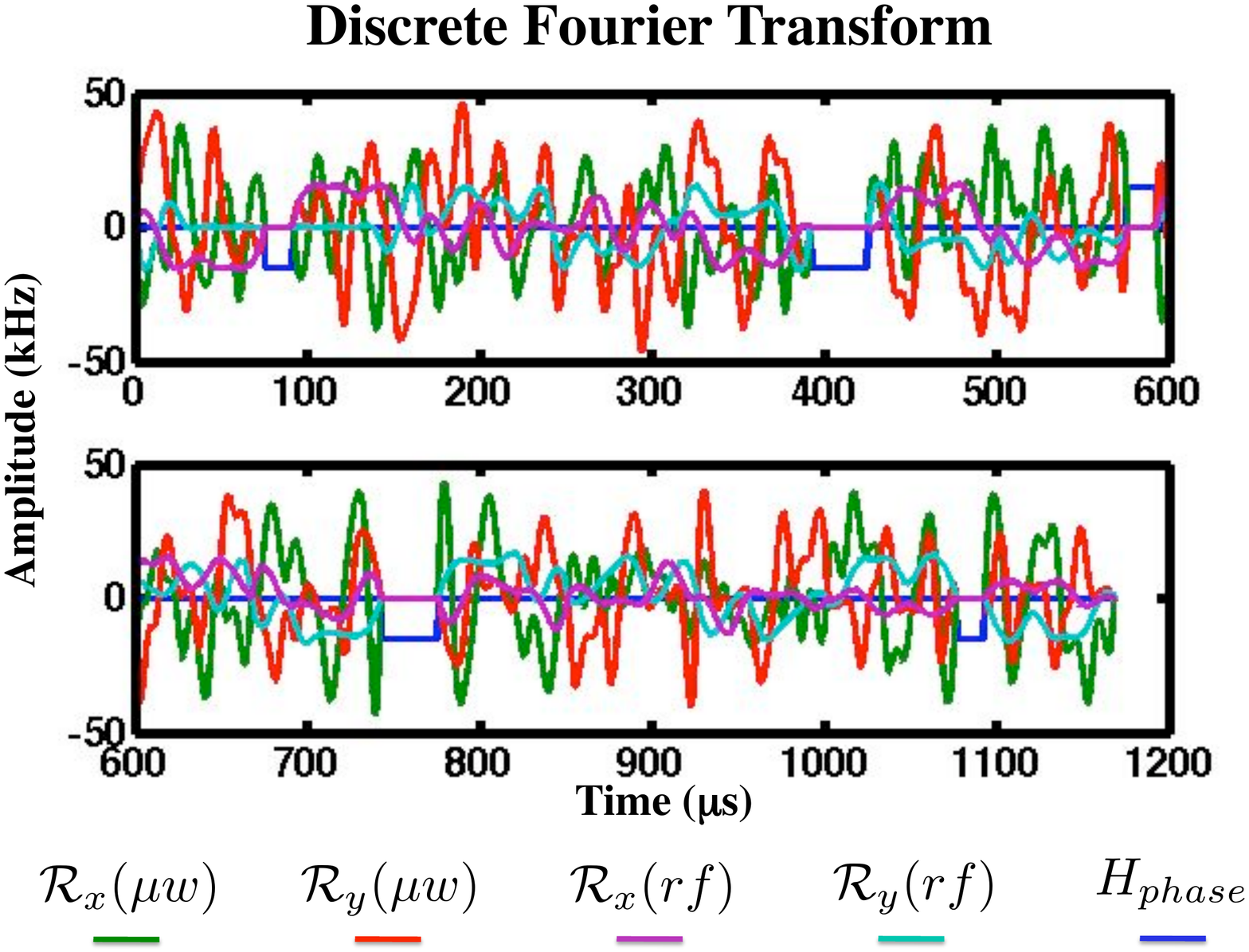}
\includegraphics[width=8cm,clip]{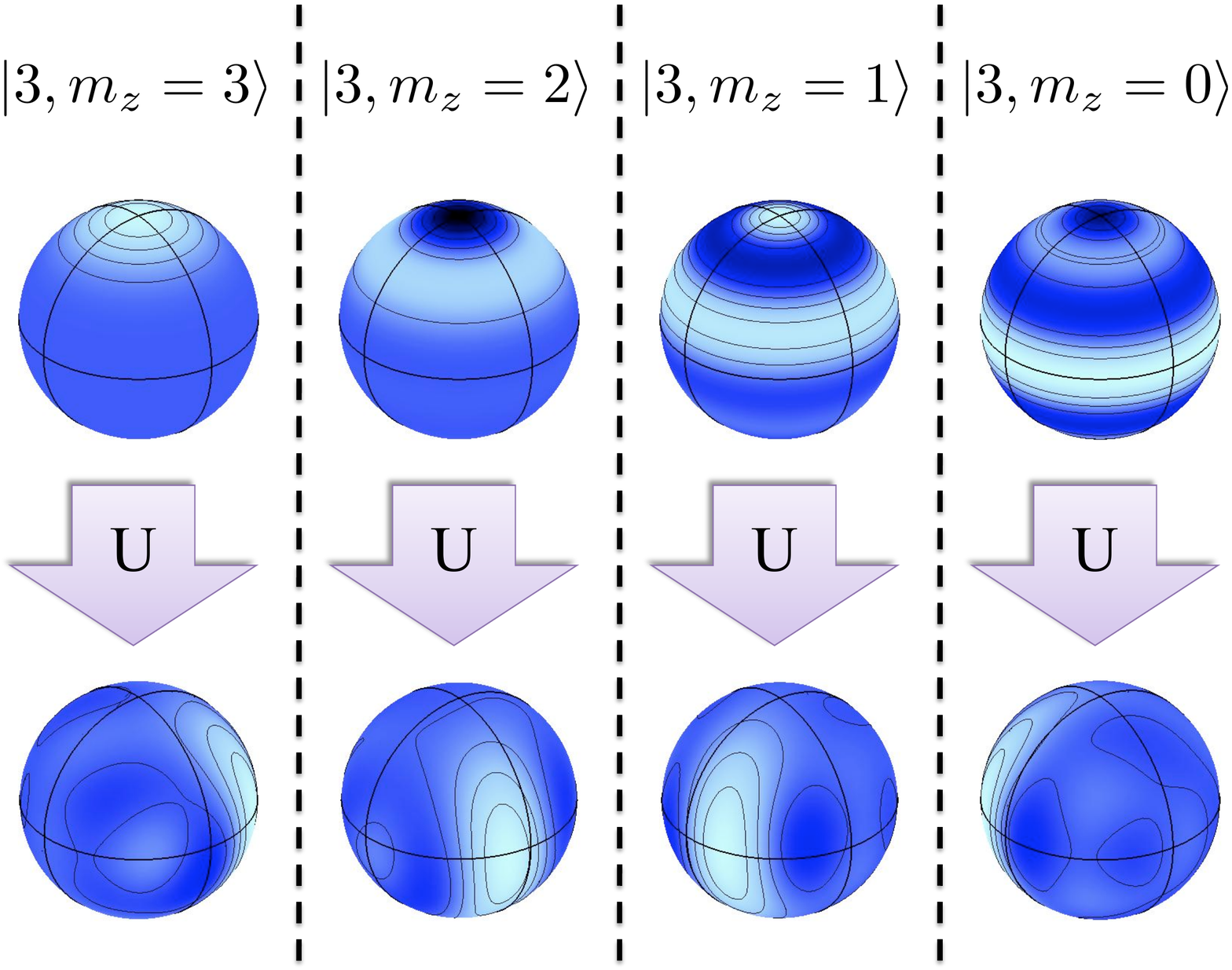}
\caption{Optimized control fields for implementing the 7-dimensional Fourier transform on the $F=3$ hyperfine manifold in $^{133}$Cs.  The duration of the pulse is less than 1.2 ms and yields a unitary map that has an overlap of 0.9854 with the desired target.  As an example, we show the action of the resulting unitary on the $Z$-eigenstates of angular momentum.  The conjugate variable of $F_z$ is the azimuthal angle $\phi$.  If we Fourier transform a $Z$-eigenstate, a state with a well defined value of $F_z$, we obtain a state that has a well defined value of $\phi$, a squeezed state.}
\label{F:wig}
\end{center}
\end{figure}

As an example, in Fig.~\ref{F:wig} we show the control sequence for the discrete Fourier transform.  The unitary map generated by this sequence should act to transform eigenstates of $Z$ into eigenstates of $X$ and vice versa.  We illustrate this through a Wigner function representation on sphere \cite{argarwal81}.  The $Z$ eigenstates are the standard basis of magnetic sublevels, whose Wigner functions are concentrated at discrete latitudes on the sphere, Fig.~\ref{F:wig}a.  Applying the control fields to each of these states yields the conjugate states, with Wigner functions shown in Fig.~\ref{F:wig}b.  These have the expected form.  They are spin squeezed states concentrated at discrete longitudes conjugate to the $Z$ eigenstates. The $Z$ and $X$ eigenstates are analogous to the number and phase eigenstates of the harmonic oscillator in infinite dimensions.

\subsection{Error-correcting a qubit embedded in a qudit}

The ability to generate unitary transformations on two-dimensional subspaces allows us to encode and manipulate a qubit in a higher dimensional Hilbert space in order to protect it from errors.  Such protection can take a passive form through the choice of a decoherence-free subspace \cite{lidar98,bacon00}, or active error correction through an encoding in a logical subspace chosen to allow for syndrome diagnosis and reversal \cite{calderbank96,aharonov97}.  Typically,  error protection schemes involve multiple subsystems (e.g. multiple physical qubits) to provide the logical subspace.  While tensor product Hilbert spaces are generally necessary to correct for all errors under reasonable noise models, for a limited error model, one can protect a qubit by encoding it an a higher dimensional qudit  \cite{gottesman01}. We consider such a protocol as an illustration of our subspace-mapping procedure.

As an example, we consider encoding a qubit in the ground-electronic hyperfine manifold of $^{133}$Cs and protecting it from dephasing due to fluctuations in external magnetic fields.  In the presence of a strong bias in the $z$-direction, the spins are most sensitive to fluctuations along that axis.  For hyperfine qubits, one solution is to choose the bias such that two magnetic sublevels see no Zeeman shift to first order in the field strength (a ``clock transition").  An alternative is to employ an active error correction protocol analogous to the familiar phase-flip code \cite{nielsen2000}.  

We take our ``physical qubit" computational basis to be the stretched states, $\ket{0}=\ket{3, 3_z}$ and $\ket{1}=\ket{4, 4_z}$, states easily prepared via optical pumping and controlled via microwave-drive rotations on the Bloch sphere.  Here we have used the shorthand labeling the two quantum numbers $\ket{F,m_z}$, and have denoted the relevant quantization axis by the subscript on the magnetic sublevel.  Such states, however, are very sensitive to dephasing by fluctuations along the bias magnetic field, and such errors are not correctable.  We choose as our encoded qubit basis stretched states along a quantization axis perpendicular to the bias ($x$-axis),  $\{\ket{\bar{0}}= \ket{3, 3_x}, \ket{\bar{1}}=\ket{3,-3_x} \}$.  Choosing this basis, a dephasing error in the $z$-direction acts to transfer probability amplitude into an orthogonal subspace.  Such errors that can be detected and reversed without loss of coherence.  

\begin{figure*}[t!]
\begin{center}
\begin{minipage}{8cm}
\includegraphics[width=7.9cm,clip]{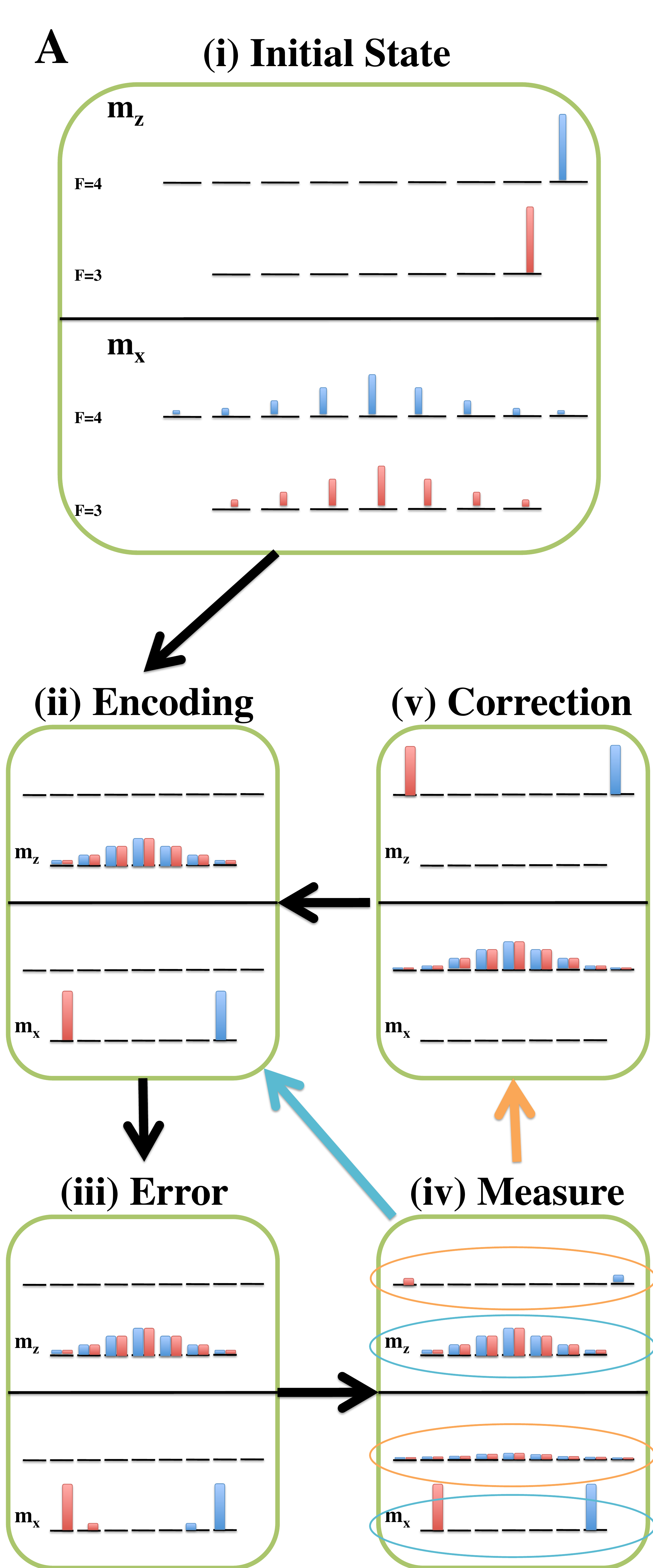}
\end{minipage}
\qquad
\begin{minipage}{8cm}
\includegraphics[width=7.9cm,clip]{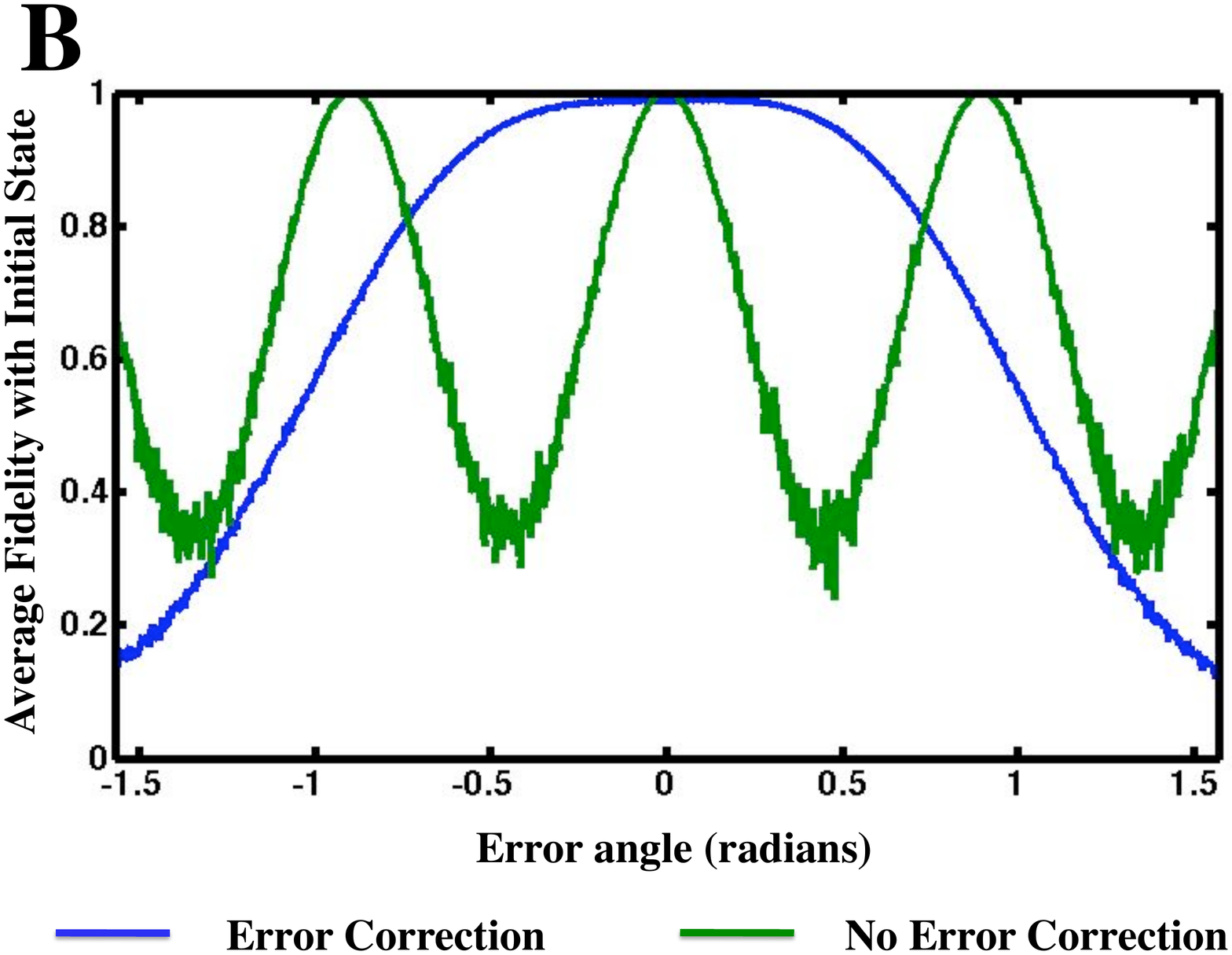}
\caption{(A)  A schematic of the error correction protocol we have designed using subspace maps.  We track the basis elements of our encoded subspace, here $\ket{0}$ is red and $\ket{1}$ is blue,  via their populations in the $x$ and $z$ bases.  (i) The initial embedded qubit we wish to protect is in a superposition of the $\ket{4,4_z}$ and  $\ket{3,3_z}$ states.  (ii)  We use subspace maps to encode the state in the basis $\ket{3,3_x}$,$\ket{3,-3_x}$.  (iii)  In this basis a small $z$-rotation shifts population into the states $\ket{3,2_x}$ and  $\ket{3,-2_x}$.  (iv)  Using subspace maps we can transfer the small population that has left the encoded space in the to states, $\ket{4,4_z}$ and $\ket{4,-4_z}$.  Now we can perform a non-demolition measurement of the total angular momentum $F$.  If $F=3$ we can be certain our state lies in the encoded subspace (ii).  If we measure $F=4$, the system is in configuration (v), which we then conditionally transform back to the encoded state.  In (B) we examine the performance of the error correction.  On the $x$-axis we have the angle of rotation in the $z$-direction due to the magnetic field error.  On the $y$-axis is the fidelity between the initial and post error states, as average over pure states drawn from the Harr measure.  The blue line shows the fidelity of the error corrected states and the green the fidelity if the state had simply stayed in the subspace $\ket{4,4_z}$,  $\ket{3,3_z}$.}
\label{F:schem}
\end{minipage}
\end{center}
\end{figure*}

Our error correction protocol works as follows (see Fig.~\ref{F:schem}). Consider an encoded qubit $\ket{\bar{\psi}} = \alpha \ket{\bar{0}} + \beta \ket{\bar{1}}$.  The error operator due to B-field fluctuations is the generator of rotations, $F_z$. Assuming a small rotation angle $2 \epsilon$, when such an error occurs, the state of our encoded qubit is mapped to
\be
e^{-2i \epsilon F_z} \ket{\bar{\psi}} \approx \ket{\bar{\psi}} + \epsilon \left( \alpha \ket{3, 2_x}+ \beta \ket{3, -2_x} \right ) . 
\ee
The error acts to spread our qubit between two orthogonal subspaces, $|m_x|=3$ and $|m_x|=2$.  To diagnose the syndrome we must measure the subspace without measuring qubit.  We can achieve this by coherently mapping the error subspace to the upper hyperfine manifold, followed by a measurement that distinguishes the two hyperfine manifolds, $F=3$ and $F=4$.   Such a coherent mapping cannot be achieved through simple microwave-driven transitions since the bias field is along the $z$-direction while the encoded states are magnetic sublevels along the $x$-direction.  We can instead use the construction of unitary operators on a subspace described in Sec.~\ref{sec:construct} to design $\pi$-rotations that take the error states to the upper manifold.  This is tricky for our implementation because our protocol only included one magnetic sublevel in the $F=4$ manifold so as to ensure proper phase imprinting.  The solution is to switch the auxiliary state in the upper manifold between two different subspace maps.  First, we consider the control system where $\ket{4,4_z}$ is our auxiliary state and perform a $\pi$-rotation that maps $\ket{3,2_x }$ to $\ket{4,4_z}$, leaving the rest of the space invariant.  Then employ control on the system where $\ket{4, -4_z}$ is the auxiliary state and map  $\ket{3,-2_x }$ to $\ket{4, -4_z}$, with the identity on the remaining space.  A QND measurement of $F$ collapses the state to the initially encoded state when the measurement result is $F=3$, or to the state $ \alpha \ket{4,4_x}+ \beta \ket{4, -4_x}$,  if we find $F=4$.  In the final step of the protocol, if an error occurred, we conditionally move the error subspace back to the encoded subspace, which can be achieved through reverse maps of the sort described above.

We simulate here the coherent steps in the error correction protocol.  These are implemented through our efficient search technique to construct subspace maps for the sequences 
\bea
\{ \ket{4,4_z}, \ket{3,3_z} \} &\rightarrow& \{ \ket{3,3_x}, \ket{3,-3_x} \}\nonumber \\
\{ \ket{3,2_x}, \ket{3,-2_x} \} &\rightarrow& \{\ket{4,4_z}, \ket{4,-4_z}\}  \nonumber \\
\{ \ket{4,4_z}, \ket{4,-4_z}\} &\rightarrow& \{ \ket{3,3_x} \ket{3,-3_x} \} \nonumber 
\eea
Each of these maps are achieved through a sequence of $SU(2)$ $\pi$-rotations on a two-dimensional subspace that leave the orthogonal subspaces invariant.  Starting from numerical searches for state preparation maps that have fidelity greater than 0.99, we obtain subspace maps with comparable fidelities.  The performance of this error correction procedure is shown in Fig.~\ref{F:schem}B.  We plot the fidelity between the initial state and the post-error-corrected state, averaged over random initial pure states of the physical qubit, versus the magnitude of the error as described by the rotation angle induced the stray magnetic field.  Even with imperfect subspace transformations the error correction protocol is significantly more robust than free evolution. Of course, like all quantum error correction protocols, we assume here that the time necessary for diagnosing the syndrome and correcting an error is sufficiently shorter than the dephasing time, so that the implementation of error correction does not increase the error probability.
                     
In practice, the most challenging step in the error correction protocol in this atomic physics example is measurement of the syndrome.  This requires addressing of individual atoms and measuring the $F$ quantum number in a manner that preserves coherence between magnetic sublevels.  In principle, this can be achieved through a QND dispersive coupling between an atom and cavity mode that induces an $F$-dependent phase shift on the field that could be detected \cite{khudaverdyan09}.  Alternatively, $F$-dependent fluorescence from a given atom would allow this code to perform ``error detection", without correction.

\section{Summary and Outlook}
In this paper we have presented a protocol for constructing unitary operators that combines the strengths of both stochastic and geometric control techniques.  By utilizing stochastic searches to construct state preparations, as opposed to stochastically searching for full unitary maps, our protocol requires computational resources that scale only polynomial with the dimension of the Hilbert space of our system.  The length of the control pulses also scales polynomially with $d$.  These stochastic search techniques place only very mild restrictions on the types of Hamiltonian controls with which our protocol is applicable.  Additionally, the controls easily generalize to the case where one wishes to control only a subspace of a larger Hilbert space.  For subspace control, the number of searches required scales as the dimension of the subspace, not as that of the embedding Hilbert space.
 
Hybrid stochastic/geometric control schemes yield a very promising path towards unitary control sequences that balance broad applicability with ease of implementation \cite{luy05}.  The most restrictive element of our protocol is the requirement that we can impart a desired phase on {\em one and only one} state (a $U(2)$ operation).  A much less restrictive procedure is to employ a control Hamiltonian that acts in imprint a relative phase between {\em two states} in the Hilbert space (an $SU(2)$ operation).  This type of operation could be implemented through, e.g., the microwave controls described in Sec.~ \ref{sec:examples}.  As a generalization of our protocol, eigenstates would be mapped pairwise to two chosen fiducial states where an external field generates the desired phase difference.  The difficulty with this approach is that we require stochastic searches for a control waveforms that maps a 2D subspace in one step, rather than than our two-step procedure which maps each basis vector separately.  The topology of the control landscape for such waveforms and complexity of such a stochastic search are not known, though we expect this to be polynomial in $d$.

While we have primarily emphasized here an exponential speedup in the search for control waveforms that generate unitary maps, a constructive protocol brings additional possible advantages.  By exploiting the geometry of a problem, we can engineer robustness more easily than in a stochastic setting.  For example, the microwave and rf-controls discussed in Sec.~ \ref{sec:examples} consist of representations of $SU(2)$ rotations in different subspaces of the Hilbert space. There are well known composite pulse techniques that implement rotations $SU(2)$ that are robust to errors in the individual pulse amplitudes and detunings \cite{vandersypen04,kobzar05}.  In future work we will explore protocols that import these methods in order to efficiently search for and  implement robust $SU(d)$ transformations.     

This research was supported by NSF Grants No. PHY-0653599 and No. PHY-0653631, ONR Grant No. N00014-05-1-420, and IARPA Grant No. DAAD19-13-R-0011.

\bibliography{Constructing_Unitaries}

\end{document}